\documentclass[11pt]{article}

\usepackage[preprint]{acl}

\usepackage{times}
\usepackage{latexsym}

\usepackage[T1]{fontenc}

\usepackage[utf8]{inputenc}

\usepackage{microtype}

\usepackage{inconsolata}
\usepackage{enumitem}

\usepackage{graphicx}

\usepackage{tcolorbox}
\usepackage[ruled,vlined]{algorithm2e}
\usepackage{url}            
\usepackage{booktabs}       
\usepackage{amsfonts}       
\usepackage{nicefrac}       
\usepackage{microtype}      
\usepackage{xcolor}         
\newcommand{\sd}[1]{\textcolor{gray}{\scriptsize$\,\pm#1$}}
\usepackage{multirow}
\usepackage{amsmath, amsthm}

\usepackage{float}

\title{Contextual Relevance and Adaptive Sampling for LLM-Based Document Reranking}

\author{
    \textbf{Jerry Huang}\textsuperscript{1}\thanks{This work was done while Jerry was a Research Intern at NewsBreak.}, 
    \textbf{Siddarth Madala}\textsuperscript{1}, 
    \textbf{Cheng Niu}\textsuperscript{2}, \\
    \textbf{Julia Hockenmaier}\textsuperscript{1}, 
    \textbf{Tong Zhang}\textsuperscript{1} \\
    \textsuperscript{1}University of Illinois at Urbana-Champaign, 
    \textsuperscript{2}NewsBreak \\
    \texttt{\{jerry8, smadala2, juliahmr, tozhang\}@illinois.edu} \\
    \texttt{cheng.niu@newsbreak.com}
}

\begin{document}
\maketitle

\begin{abstract}
Reranking algorithms have made progress in improving document retrieval quality by efficiently aggregating relevance judgments generated by large language models (LLMs). However, identifying relevant documents for queries that require in-depth reasoning remains a major challenge. Reasoning‐intensive queries often exhibit multifaceted information needs and nuanced interpretations, rendering document relevance inherently context dependent. To address this, we propose \emph{contextual relevance}, which we define as the probability that a document is relevant to a given query, marginalized over the distribution of different reranking contexts it may appear in (i.e., the set of candidate documents it is ranked alongside \emph{and} the order in which the documents are presented to a reranking model). While prior works have studied methods to mitigate the positional bias LLMs exhibit by accounting for the ordering of documents, we empirically find that the compositions of these batches also plays an important role in reranking performance. To efficiently estimate contextual relevance, we propose \textbf{TS-SetRank}, a sampling-based, uncertainty-aware reranking algorithm. Empirically, TS-SetRank improves nDCG@10 over retrieval and reranking baselines by 15–25\% on BRIGHT and 6–21\% on BEIR, highlighting the importance of modeling relevance as context-dependent.
\end{abstract}

\section{Introduction}
\label{sec:intro}

\begin{figure*}[t]
  \centering
  \includegraphics[width=\textwidth]{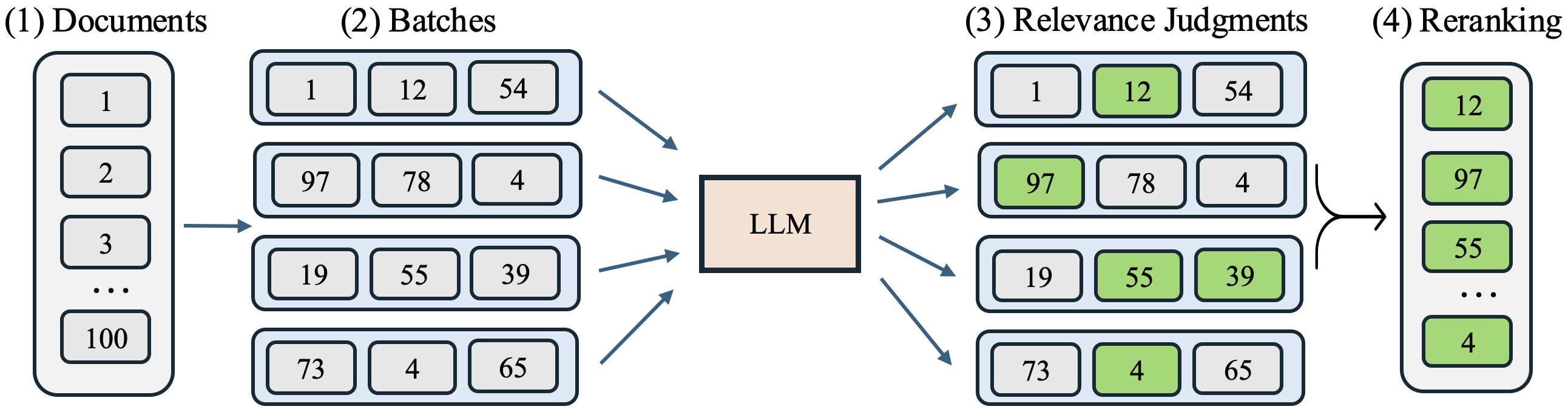}
  \caption{An overview of the Setwise prompting approach for document reranking. Here, we draw batches of size three from the initial list of retrieved documents and prompt an LLM to identify all relevant documents in each batch (highlighted in green). The final reranking list is formed by aggregating all relevance judgments across batches.}
  \label{fig:contextual-relevance}
\end{figure*}

Large language models (LLMs) have shown strong performance in zero-shot document reranking~\cite{chen2025attention, 10.1145/3626772.3657951}, with the Setwise prompting approach~\cite{Zhuang_2024_setwise} offering a favorable trade-off between the number of LLM inference calls necessary for reranking and final reranking quality. As illustrated in Figure~\ref{fig:contextual-relevance}, in this approach, smaller subsets or batches of documents are drawn from the initial set of retrieved documents and presented to an LLM trained for reranking, which in turn generates per-document binary judgments of relevance to a given query. Final rankings are then constructed by aggregating the model’s relevance judgments across batches using aggregation or well-known sorting algorithms such as bubble sort or heap sort. 

Final reranking quality is thus determined by the ability of a reranking model to identify all relevant documents in each batch, while constrained by a fixed inference budget. While past work has investigated the positional bias LLMs exhibit when making relevance judgments~\cite{tang-etal-2024-found, vardasbi2025adaptiverepetitionmitigatingposition, li-etal-2024-split}, we find that the context in which documents are presented to LLMs also impacts their ability to identify all relevant documents. This phenomenon is particularly evident in queries that require deeper levels of reasoning, have multi-faced information needs, and/or contain nuanced interpretations~\cite{zeng2024llmrankfusionmitigatingintrinsicinconsistency, Gienapp_2022}. As highlighted by BRIGHT~\cite{su2025brightrealisticchallengingbenchmark}, a benchmark for reasoning-intensive retrieval, both first-stage retrievers and reranking models struggle to retrieve the set of all relevant documents in such cases.

As a descriptive example, consider the following prompt:
\begin{quote}
\emph{Judge the relevance of the following passages to the query, 
``What is the primary ecological role of sea otters in kelp forest ecosystems?''}

\begin{enumerate}[label=(\arabic*)]
    \item \emph{Sea otters are considered a keystone species because they control sea urchin populations, which in turn helps maintain healthy kelp forests.}
    \item \emph{Unchecked sea urchin populations can devastate kelp forests, making predator control essential for ecosystem stability.}
    \item \emph{Marine mammals such as seals, whales, and dolphins play critical roles in ocean nutrient cycling.}
\end{enumerate}
\end{quote}
Passage (1) is consistently judged relevant, as it directly connects sea otters to the query. Passage (2) is also relevant but its relevance is dependent on the context of (1). Passage (3) is irrelevant. Thus, a reranking model might incorrectly judge all the documents in a batch formed by documents (2) and (3) as irrelevant. 

To capture these effects, we define \emph{contextual relevance} as the probability that a document is judged relevant given a query marginalized over all batches in which it may appear, where each batch is defined by both its contents and the ordering of its documents. Although this formulation assumes independence across documents, for tractability, marginalizing over all possible batches allows us to recover each document’s expected relevance across diverse contexts. This effectively captures contextual dependencies in expectation even when individual batch judgments are interdependent.

A natural approach to rerank a list of documents is to enumerate all possible batches in which each document may appear. However, to more efficiently estimate contextual relevance, we propose Thompson Sampling for Setwise Reranking (TS-SetRank), a two-phase Bayesian reranking algorithm that first samples document batches uniformly to collect unbiased relevance feedback, and then adaptively constructs batches using Thompson sampling~\cite{agrawal2012analysisthompsonsamplingmultiarmed}. We evaluate TS-SetRank on BRIGHT~\cite{su2025brightrealisticchallengingbenchmark} and BEIR~\cite{thakur2021beirheterogenousbenchmarkzeroshot}, two benchmarks for evaluating retrieval capabilities, and compare its Normalized Cumulative Discount Gain (nDCG@k;~\citealp{Jrvelin2002CumulatedGE}) performance against deterministic reranking and statistical retrieval baselines.


Our contributions are as follows:
\begin{itemize}
  \item We propose the notion of \textit{contextual relevance}, which takes into account both the contents and ordering of retrieved documents within each batch for Setwise reranking.
  \item We show that variance in LLM-based setwise relevance judgments arises from LLM sampling variability and contextual dependencies, and empirically quantify the share of each.
  \item We propose TS-SetRank and show that it outperforms deterministic reranking algorithms in nDCG@10 under comparable inference budgets on BRIGHT and BEIR.
\end{itemize}

\section{Estimating Contextual Relevance}
\label{sec:methods}

\subsection{Problem Formalization}

\paragraph{Objective.}
Let \(\mathcal{Q}\) denote the space of user queries and \(\mathcal{D}\) the document corpus. For each query \(q \in \mathcal{Q}\), a first-stage retriever returns an ordered list of candidate documents:
\[
D = \bigl(d_{(1)}, d_{(2)}, \dots, d_{(N)}\bigr) \subseteq \mathcal{D},
\]
where \(d_{(1)}\) is the highest-scoring document and \(N \ll |\mathcal{D}|\). The reranking objective is to select an ordered subset of \(D\) that maximizes the final reranking quality as measured by a specified retrieval metric such as nDCG@k.

\paragraph{Reranking Model.} 
A setwise reranking model is an LLM that takes a query $q$ and a small batch of documents $S = (d_1,\dots,d_b)$ where $b\ll N$, and makes a binary relevance judgment for each document.

\paragraph{Contextual Relevance.} 
We define the \emph{contextual relevance} of a document $d_i$ with respect to a query $q$ as its probability of being judged relevant, marginalized over all batches in which it may appear:
\[
  \theta_{i,q} = \mathbb{E}_{S \sim D_b(d_i)}
  \Bigl[ \Pr\bigl(d_i \text{ is judged relevant} \mid q, S\bigr) \Bigr]
\]
where $D_b(d_i)$ denotes the distribution over size-$b$ batches that include $d_i$, and $D_b(D)$ denotes the distribution over all size-$b$ batches drawn from $D$. We use $\theta_i$ interchangeably with $\theta_{i,q}$, as the query $q$ is fixed within each reranking task.

\paragraph{Modeling Assumptions.} 
This formulation assumes conditional independence across documents: each document’s judgments are modeled as independent draws following a Bernoulli distribution with mean $\theta_{i,q}$. While this simplification is reasonable when documents do not exhibit strong dependencies, more complex models (e.g., structured bandits or Plackett--Luce) would be required in domains such as multi-hop question answering~\cite{mavi2024mhqa} or citation retrieval~\cite{Qian2024OnTC}, where relevance depends on prerequisite evidence. Benchmarks like BRIGHT and BEIR do not exhibit such dependencies, making this assumption appropriate in our setting.

\subsection{Bayesian Reranking}
We model inference-time reranking as a combinatorial semi-bandit problem with a fixed budget of \(T\) rounds. At each round \(t = 1, \dots, T\), the model selects a batch \(S_t \sim D_b(D)\) and queries the reranking model for binary relevance feedback. We maintain independent Beta-Bernoulli posteriors for each document \(d_i \in D\), initialized with $\alpha_i = \beta_i = 1$, so that $\theta_i \sim \mathrm{Beta}(\alpha_i, \beta_i)$ represents our uncertainty over $d_i$’s contextual relevance. 

At each round, we perform:
\begin{enumerate}
\item \textbf{Batch Selection.}
  We sample a batch
  $S_t \;\sim \;D_b(D)$ according to our policy
  (e.g.\ uniform exploration or Thompson sampling, defined in Section~\ref{sec:ts-setrank}).

\item \textbf{Setwise Feedback.}
  For each batch $S_t$, the reranker makes a binary relevance decision for each 
  document. $R(q, S_t)$ denotes the subset of documents judged relevant, and $r_t(d_i) = 1$ if $d_i \in R(q, S_t)$ and $0$ otherwise.

\item \textbf{Posterior Update.}
  We update the Beta posterior for each $d_i \in S_t$ following:
  \[
  \alpha_i \leftarrow \alpha_i + r_t(d_i),
  \quad
  \beta_i \leftarrow \beta_i + (1 - r_t(d_i)),
  \]
  where the posterior mean
  \[
  \hat\theta_i = \frac{\alpha_i}{\alpha_i + \beta_i}
  \quad \forall\, d_i \in D
  \]
  serves as our current estimate of $d_i$’s contextual relevance.

\end{enumerate}
The final ranking is computed by sorting the candidate set $D$ in descending order by the posterior means $\hat\theta_i$.

\subsection{TS-SetRank}
\label{sec:ts-setrank}
We now introduce Thompson Sampling for Setwise Reranking (TS-SetRank) as outlined in Algorithm~\ref{alg:ts-setrank}, a two-phase algorithm that combines uniform sampling with Thompson sampling to guide batch selection. We maintain independent Beta--Bernoulli posteriors for each document $d_i$’s relevance and adaptively allocate queries using posterior sampling, aiming to improve ranking quality under a fixed inference budget $T$.

\begin{algorithm}[t]
\caption{TS-SetRank}
\KwIn{First-stage retrieval results $D = \{d_1, \dots, d_N\}$, query $q$, budget $T$, batch size $b$, exploration rounds $T_f$, reranking model $M$}
\KwOut{Final reranked list of documents}

\ForEach{$d_i \in D$}{
  Initialize posterior: $\alpha_i \gets 1$, $\beta_i \gets 1$
}

\For{$t = 1$ \KwTo $T$}{
  \uIf{$t \leq T_f$}{
    Sample batch $S_t \sim \mathrm{Uniform}\!\bigl(D_b(D)\bigr)$
  }
  \Else{
    \ForEach{$d_i \in D$}{
      Draw $\tilde\theta_i \sim \mathrm{Beta}(\alpha_i, \beta_i)$
    }
    Let $S_t \gets \{\,d_i \in D : \tilde\theta_i \text{ is among the top-}b \,\}$
  }
  
  Obtain feedback: $R_t := R(q, S_t)$ from reranking model $M$

  \ForEach{$d_i \in S_t$}{
    Observe $r_t(d_i) \in \{0,1\}$ \\
    Update posterior: $\alpha_i \gets \alpha_i + r_t(d_i)$, 
    $\beta_i \gets \beta_i + (1 - r_t(d_i))$
  }
}

\Return documents sorted by posterior means $\hat\theta_i = \alpha_i / (\alpha_i + \beta_i)$ in descending order.
\label{alg:ts-setrank}
\end{algorithm}

TS-SetRank proceeds in two phases: $T_f$ rounds of uniform sampling followed by $T-T_f$ rounds of adaptive sampling.

\begin{itemize}
    \item \textbf{Phase I: Uniform Sampling.} For rounds $t = 1, \ldots, T_f$, we sample batches $S_t$ uniformly at random from ${D}_b(D)$. This provides sufficient initial observations for posterior inference. While one could initialize from a first-stage retriever ranking, we opt for uniform sampling given the relatively weak performance of first-stage retrievers on our chosen benchmarks.
    
    \item \textbf{Phase II: Adaptive Sampling.} For $t = T_f + 1, \ldots, T$, we draw samples $\tilde\theta_i \sim \mathrm{Beta}(\alpha_i, \beta_i)$ from each document’s posterior and form the batch $S_t$ by selecting the $b$ documents with the highest sampled values. Thompson sampling balances exploration and exploitation by sampling from the posterior: documents with higher estimated relevance are more likely to be chosen, while uncertain documents still have a chance of being selected to discover overlooked relevant items~\cite{agrawal2012analysisthompsonsamplingmultiarmed}.

\end{itemize}

\subsection{Theoretical Guarantees}
\label{sec:theory}

We summarize the theoretical properties of TS-SetRank based on established analyses of Thompson sampling in stochastic semi-bandit settings.

\paragraph{Sublinear Regret.}
Reranking quality in our experiments is measured by nDCG@10. For analytical tractability, we instead consider a linear \emph{surrogate objective} in which each document $d_i$ is treated as an independent Bernoulli arm with mean contextual relevance~$\theta_i$. At each round, the model selects a batch of $b$ documents from the pool of $N$ retrieved documents and observes binary relevance feedback. The surrogate reward is defined as the sum of these feedback signals, representing the number of relevant documents retrieved in the batch.

We note, however, that maximizing this surrogate encourages selection of documents with high expected relevance but does not directly optimize the position-weighted nDCG@10 objective. Yet, in practice, improving expected relevance tends to increase nDCG@10 empirically.

Under standard assumptions of independent arms, bounded rewards, and sufficient exploration, Thompson sampling in stochastic semi-bandit settings achieves sublinear cumulative regret~\cite{agrawal2012analysisthompsonsamplingmultiarmed, pmlr-v28-chen13a, pmlr-v80-wang18a}:
\[
\mathbb{E}[\mathcal{R}(T)] = \tilde O\!\bigl(\sqrt{bNT}\bigr),
\]
where $b$ is the batch size, $N$ the number of retrieved documents, $T$ the number of rounds, and $\mathcal{R}(T)$ denotes the cumulative surrogate regret. Since TS-SetRank follows the same structure, it inherits this sublinear regret bound for the surrogate estimation task.

\paragraph{Posterior Consistency.}
For each document $d_i$, feedback observations are Bernoulli-distributed with mean~$\theta_i$. Under the standard Beta--Bernoulli update, the posterior mean
$\hat{\theta}_i=\alpha_i/(\alpha_i+\beta_i)$ converges almost surely to the true mean~$\theta_i$ whenever the document is sampled infinitely often, by the strong law of large numbers and Beta--Bernoulli conjugacy.

\paragraph{Uniform Exploration.}
A non-adaptive policy such as uniform sampling does not exploit accumulated feedback and therefore incurs linear regret, $\mathcal{R}(T)=\Theta(T)$~\citep{auer2002finite}.
This contrast highlights the benefit of adaptive exploration in reducing long-run cumulative error.

\section{Experiments}
\label{experiments}
To evaluate the effectiveness of our framework for modeling contextual
relevance and the performance of TS-SetRank, we organize our experiments around
the following research questions:

\begin{itemize}
  \item[\textbf{RQ1.}] To what extent is variability in LLM-based relevance judgments attributable to LLM sampling variability versus changes in the surrounding document context?
  
  \item[\textbf{RQ2.}] How does TS-SetRank compare with uniform sampling,
  deterministic reranking methods, and standard retrieval baselines
  (e.g., BM25) in terms of nDCG@10 under a fixed inference budget?

  \item[\textbf{RQ3.}] How does uniform sampling improve reranking performance as the inference budget grows, and when does it reach convergence?
  
\end{itemize}

\subsection{Experimental Setup}

\paragraph{Reranking Model.}
We train the Setwise reranking model used for all our experiments by fine-tuning the Qwen2.5-7B-Instruct base model~\cite{qwen2025qwen25technicalreport} on 25{,}000 training samples from the MS MARCO v2.1 dataset~\cite{bajaj2018msmarco}. Post-training is performed using reinforcement learning with verifiable rewards (RLVR) via Group Relative Policy Optimization (GRPO;~\citealp{shao2024deepseekmathpushinglimitsmathematical}). Following prior work on reasoning-enhanced rerankers~\cite{zhuang2025rankr1enhancingreasoningllmbased, weller2025rank1}, our model is trained to leverage test-time compute by first producing an explicit reasoning trace enclosed in predefined reasoning tags before emitting the set of passages it deems relevant.

\paragraph{Reward Function.}
We define two reward components: a formatting reward and a correctness-based reward. The formatting reward ensures that the model produces outputs enclosed within valid tags (e.g. <answer>...</answer>), which are required for evaluation. Specifically, the model receives a reward of $0.5$ if both the opening and closing tags are present and correctly formatted, and $0$ otherwise. 

The correctness reward measures agreement between the predicted and gold document sets using the $F_\beta$ metric, computed only when valid <answer> tags are detected; otherwise, it is assigned $0$. We use $\beta = 2$ to emphasize recall:
\[
F_\beta = (1 + \beta^2) \cdot \frac{\text{precision} \cdot \text{recall}}{\beta^2 \cdot \text{precision} + \text{recall}}.
\]
The total reward is the sum of the formatting and correctness components.

\paragraph{Training Details.}
All training samples contain a query and batches of size 10, each containing between 1 and 7 relevant documents with the positions of the relevant documents randomly determined for each training sample. All fine-tuning is conducted on H100 GPUs. Additional implementation details, including the specific prompt templates and hyperparameters used are provided in Appendix~\ref{app:training}.

\paragraph{Datasets \& Retrieval}  
We benchmark our model and reranking algorithms on  BRIGHT~\cite{su2025brightrealisticchallengingbenchmark} and BEIR~\cite{thakur2021beirheterogenousbenchmarkzeroshot}, two widely used benchmarks for evaluating retrieval and reranking quality. Following prior work, we use BM25~\cite{bm25, bm25s} as the first-stage retriever and use the top-100 retrieved documents for reranking. We set the batch size to $10$ for sampling-based reranking algorithms and set the total inference budget to $T=100$ as is common in prior works. For BRIGHT, we use the benchmark provided GPT-4o rewritten queries to improve the quality of our first-stage retrieval results. For BEIR, we benchmark on the datasets containing fewer than 2,000 queries and sample 100 queries per dataset for evaluation.

\begin{table*}[t]
\centering
\small
\begin{tabular}{@{}lccc@{\hskip 2em}ccc@{}}
\toprule
 & \multicolumn{3}{c}{\textbf{Batch size $b=2$}}
 & \multicolumn{3}{c}{\textbf{Batch size $b=10$}} \\
\cmidrule(lr){2-4} \cmidrule(lr){5-7}
\textbf{Regime} & Intrinsic & Positional & Total 
                & Intrinsic & Positional & Total \\
\midrule
\textbf{Accuracy} 
  & 0.27 & 0.26 & 0.26
  & 0.28 & 0.28 & 0.27 \\

\textbf{Variance} 
  & 0.063 
  & 0.076 
  & 0.083 
  & 0.062 
  & 0.103 
  & 0.113 \\
\bottomrule
\end{tabular}
\caption{
Mean accuracy and per-query variance of LLM judgments on BRIGHT for batch sizes $2$ and $10$ under three regimes: 
\emph{Intrinsic} (fixed batch documents, fixed order), 
\emph{Positional} (fixed batch documents, shuffled order), 
and \emph{Total} (resampled documents, shuffled order). 
Variance is computed per query across repeated trials and then averaged. 
}
\label{tab:reranking_model_variability}
\end{table*}

\paragraph{Baselines \& Variants}
\begin{itemize}
  \item \textbf{BM25:} A statistical retrieval method based on term frequency and inverse document frequency, widely used as a first-stage retriever. BM25 serves as a strong non-neural baseline.
  
  \item \textbf{Heapify:} A deterministic setwise reranking algorithm that performs a sequence of binary heap comparisons over document triplets, requiring a total of \(O(N \log N)\) LLM inference calls. \citet{Zhuang_2024_setwise} show that Heapify achieves strong performance under a fixed budget.
  
  \item \textbf{TS-SetRank (Uniform):} A variant of TS-SetRank that performs 100 steps of uniform sampling with no adaptive Thompson sampling phase. This corresponds to the case TS(\(100/0\)), and serves as a non-adaptive baseline for comparison.
  
  \item \textbf{TS-SetRank (\(X/100{-}X\)):} Our full TS-SetRank algorithm, parameterized by \(X\) steps of uniform sampling followed by \(100{-}X\) steps of Thompson sampling. Varying \(X\) allows us to study the exploration–exploitation trade-off and assess how much adaptive sampling improves performance.
\end{itemize}

\paragraph{Metrics}
We report nDCG@10 after \(T = 100\) reranking steps, along with intermediate performance snapshots to visualize the progression of reranking quality. All results are reported as mean ± one standard deviation across three random seeds. 

\subsection{RQ1: Reranking Model Variability}
We first empirically quantify how much of the variability in LLM-based relevance judgments can be attributed to intrinsic model stochasticity (token sampling)  versus contextual factors such as batch composition and document order.

\paragraph{Experimental Setup.} 
For each query in BRIGHT, we randomly select one relevant document $d^+$ from the ground-truth set and construct 30 batches of size $b$. Each batch contains $d^+$ alongside $b{-}1$ other documents sampled from the top-100 BM25 candidates. For each batch, accuracy is recorded as $1$ if $d^+$ is judged relevant and $0$ otherwise. This setup abstracts away from ranking metrics such as nDCG and focuses exclusively on per-document judgment variability at the batch level.

We evaluate three regimes:
\begin{enumerate}
  \item \textbf{Intrinsic.} All 30 batches we construct are identical in both composition and order, so variability reflects only randomness introduced by token sampling.
  \item \textbf{Positional.} Batch composition is fixed, but document order is shuffled. Variability here includes both intrinsic randomness and any additional sensitivity to order.
  \item \textbf{Total.} Distractors are resampled each time and order randomized, so variability includes intrinsic, positional, and compositional effects.
\end{enumerate}

\paragraph{Batch Size Effects.}  
Table~\ref{tab:reranking_model_variability} reports mean accuracy and per-query variance under each regime for batch sizes $2$ and $10$. We observe that accuracy improves modestly with larger batch sizes, but overall variability also increases. The intrinsic baseline remains stable across batch sizes, suggesting that the additional variability arises from contextual factors, particularly to document order when $b$ is larger.

Because the regimes build on one another, we can quantify the contributions of each by computing their differences. This shows how much additional variability is attributable to ordering effects beyond intrinsic variability, and how much more variability is added when we also consider compositional effects. From Table~\ref{tab:reranking_model_variability}, we find that contextual factors explain about 25\% of the total variability for size-2 batches and 45\% for size-10 batches. Most of the increase comes from sensitivity to order (16\% and 36\%, respectively), while compositional effects remain fairly steady at around 9\%. These results serve only as an \textit{empirical heuristic} as the sources of variance are not completely independent.

\begin{figure}[t]
  \centering
  \includegraphics[width=1\linewidth]{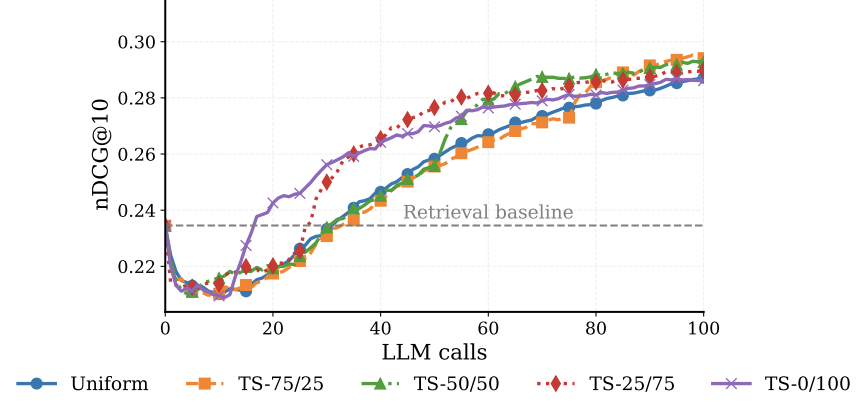}
  \caption{
TS-SetRank variants achieve faster gains under smaller inference budgets, particularly when using fewer exploration rounds. The initial dip in reranking quality reflects noisy posterior estimates early in sampling, but performance quickly surpasses the retrieval baseline as more rounds are completed.
}
  \label{fig:summary_plots}
\end{figure}

\begin{table*}[t]
\centering
\footnotesize
\resizebox{\linewidth}{!}{%
  \begin{tabular}{lcccccccc}
    \toprule
    Benchmark & Snapshot \(t\) & BM25s      & Heapify        & Uniform         & TS-75/25         & TS-50/50         & TS-25/75         & TS-0/100         \\
    \midrule
    \multirow{2}{*}{BRIGHT}
               & 100 & 0.235      & 0.2560 & 0.287\sd{0.003} & \textbf{0.294}\sd{0.002} & 0.293\sd{0.003} & 0.290\sd{0.004} & 0.286\sd{0.002} \\
               & 50  & \textemdash   & \textemdash     & 0.258\sd{0.003} & 0.255\sd{0.004}\(\dagger\) & 0.255\sd{0.003}\(\dagger\) & \textbf{0.276}\sd{0.006} & 0.270\sd{0.003} \\
    \addlinespace
    \multirow{2}{*}{BEIR}
               & 100 & 0.357      & 0.4080 & 0.421\sd{0.003} & 0.429\sd{0.002}         & 0.424\sd{0.004} & \textbf{0.431}\sd{0.010} & 0.428\sd{0.005} \\
               & 50  & \textemdash  & \textemdash     & 0.393\sd{0.000} & 0.401\sd{0.009}\(\dagger\) & 0.395\sd{0.002}\(\dagger\) & \textbf{0.417}\sd{0.001} & \textbf{0.417}\sd{0.005} \\
    \bottomrule
  \end{tabular}%
}
\caption{
nDCG@10 at snapshots $t$ at 50 and 100 on BRIGHT and BEIR. 
$\text{TS-}\alpha/(100-\alpha)$ denotes TS-SetRank with \(\alpha\) uniform exploration steps and \(100{-}\alpha\) Thompson sampling steps. Uniform corresponds to TS-SetRank with 100 rounds exploration. 
\(\dagger\) indicates snapshots before Thompson sampling begins. Full topic-level results for both benchmarks are provided in Appendix~\ref{app:bright_results} and~\ref{app:beir_results}.
}
\label{tab:combined_averages}
\end{table*}

\subsection{RQ2: Comparison of Reranking Algorithms}
We evaluate TS-SetRank against deterministic and non-adaptive reranking baselines under a fixed inference budget of \(T{=}100\). Table~\ref{tab:combined_averages} reports nDCG@10 at two checkpoints (\(t{=}50\) and \(100\)) on BRIGHT and BEIR.

At \(t{=}100\), Uniform achieves strong performance, reflecting the benefit of averaging judgments across reranking contexts. TS-SetRank attains comparable or slightly better final performance across different exploration–exploitation splits, with dataset-specific variation: on BRIGHT, TS-75/25 achieves the best score (0.294), while on BEIR, TS-25/75 slightly outperforms other variants. These results suggest that adaptive sampling is at least as effective as uniform sampling in the long run.

The advantage of TS-SetRank is most evident under smaller budgets. At the halfway point (\(t{=}50\)), illustrated in Figure~\ref{fig:summary_plots} and reported in Table~\ref{tab:combined_averages}, TS-SetRank variants with more exploitation (e.g., TS-25/75 and TS-0/100) outperform Uniform. On BRIGHT, TS-25/75 improves upon Uniform by 1.8 points, and on BEIR, TS-25/75 and TS-0/100 reach 0.417 nDCG@10 compared to 0.393 for Uniform. This indicates that TS-SetRank can more effectively allocate its limited inference budget by focusing on promising candidates earlier.

Heapify underperforms across both datasets due to its reliance on potentially noisy pairwise comparisons. In such regimes, relevant documents often fail to consistently accumulate wins to rise to the top of the heap. This limitation aligns with noisy sorting theory, which shows that reliable rankings under stochastic comparisons require repeated resampling and aggregation~\cite{braverman2007noisysortingresampling}. 

\subsection{RQ3: Convergence of Uniform Sampling}

To investigate how uniform sampling behaves as the inference budget increases, we examine its convergence characteristics under an unconstrained evaluation setting. Unlike the adaptive TS-SetRank variants, uniform sampling performs all reranking steps by drawing document batches uniformly at random, without leveraging feedback from previous rounds. This setup allows us to isolate the effect of repeated contextual averaging on final ranking quality.

\begin{figure}[t]
  \centering
  \includegraphics[width=0.8\linewidth]{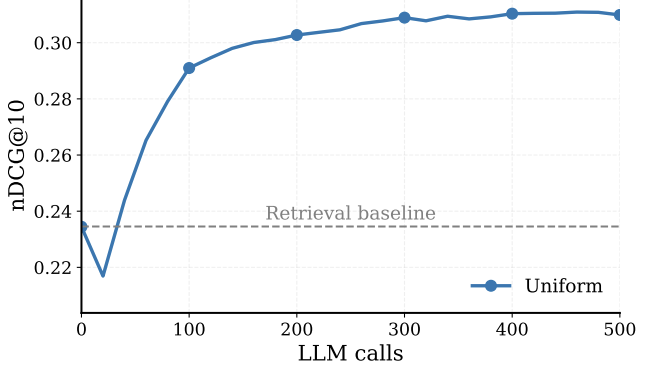}
  \caption{
nDCG@10 vs.\ LLM calls on BRIGHT.
Uniform sampling converges after $\approx300$ inference calls, showing diminishing returns.}
  \label{fig:uniform}
\end{figure}

As shown in Figure~\ref{fig:uniform}, nDCG@10 improves rapidly during the initial 100--200 inference calls, followed by progressively smaller gains beyond 300 calls.  Empirically, convergence occurs after approximately 300 inference calls, at which point additional sampling yields negligible improvements in nDCG@10.

\subsection{Inference Costs \& Parallelism}
\label{inference_costs}
Both TS-SetRank and Uniform sampling operate under a fixed inference budget of 50 or 100 inference calls per query, each evaluating batches of 10 documents. In contrast, Heapify follows a binary comparison tree: it performs 50 initial comparisons (one per internal node for $n{=}100$) and up to 7 additional comparisons per extraction during its top-10 selection phase, yielding an average of 110 LLM inference calls per query in our experiments. Fully reranking all 100 documents requires up to 743 comparisons in the worst case. While individual TS-SetRank calls process more tokens due to larger batch sizes, amortizing computation across multiple documents substantially improves throughput. Empirically, our results in Table~\ref{tab:main_results_app} shows that at the snapshot of $t{=}50$, TS-SetRank achieves stronger performance while remaining more efficient when compared to Heapify.

Uniform sampling trivially parallelizes across queries and document batches, achieving linear scaling with the number of available workers. By contrast, Heapify exposes structured but depth-dependent parallelism: its binary comparison tree allows concurrent evaluation of all comparisons within a level once their descendants have been resolved, yielding $O(n)$ total work and an ideal span of $O(\log n)$. This design enables high concurrency near the leaves but diminishing efficiency toward the root, where the active frontier narrows and synchronization costs dominate. 

\paragraph{Extensions to Delayed Feedback.}
Phase II of TS-SetRank is not parallelizable due to the reliance of feedback from previous rounds. To increase the parallelism of TS-SetRank, we propose a throughput optimized variant, TS-SetRank-T (see Appendix~\ref{app:ts_setrank_t}), which extends TS-SetRank by deferring parameter updates within each phase while preserving Thompson sampling semantics. Under mild regularity assumptions, such delayed-update schemes maintain sublinear regret in the combinatorial semi-bandit setting with delayed feedback~\cite{joulani2013onlinelearningdelayedfeedback, karbasi2021parallelizing}, enabling more efficient contextual relevance estimation when more workers are available.

\section{Related Work}
\label{sec:background}

\subsection{Reranking Paradigms}
The four dominant prompting approaches for using LLMs in zero-shot document reranking include Pointwise~\cite{sachan2023improvingpassageretrievalzeroshot}, Pairwise~\cite{qin2024largelanguagemodelseffective}, Listwise~\cite{ma2023zeroshotlistwisedocumentreranking, pradeep2023rankvicunazeroshotlistwisedocument, pradeep2023rankzephyr}, and Setwise~\cite{Zhuang_2024_setwise} prompting. Respectively, each of these prompting approaches employs an LLM to generate: (1) a binary judgment of relevance for a single document, (2) a preference judgment for a pair of documents, (3) an ordered list of relevance scores given a list of documents, and (4) a subset of relevant documents from a candidate set. In this work, we focus on improving Setwise reranking models, which have been shown to balance efficiency and accuracy by allowing for multiple pairwise comparisons to be batched within a single Setwise prompt. Moreover, because most retrieval datasets are designed for Pointwise or Setwise evaluation, training effective listwise reranking models remains challenging~\cite{zhang2023rankwithoutgptbuildinggptindependentlistwise}. Other works have also studied the performance of training reasoning language models (RLMs) specifically for the task of document reranking and have observed improved performance on reasoning-based document retrieval tasks~\cite{zhuang2025rankr1enhancingreasoningllmbased, weller2025rank1}.

\subsection{Non-Transitivity of LLM Judgments}
The transitivity of pairwise judgments has been well-studied in the field of information retrieval, with empirical studies showing that this property holds reliably only when the judgment model is well-trained~\cite{Hui2017TransitivityTC, xu2025investigatingnontransitivityllmasajudge}. In LLM-based reranking, studies have found that using smaller models or benchmarking on complex or subjective tasks correlate with higher intransitivity~\cite{liu2025aligninghumanjudgementrole}. Reasoning-intensive queries also exhibit intransitive judgments as they often involve nuanced forms of relevance (i.e., multiple valid ways a document can relate to a query). To mitigate intransitivity and improve reranking, prior work has proposed methods such as aggregating judgments from multiple reranking models~\cite{zeng2024llmrankfusionmitigatingintrinsicinconsistency}, subsampling a larger set of pairwise comparisons~\cite{Gienapp_2022}, using permutation‐based self‐consistency to marginalize positional bias~\cite{tang-etal-2024-found}, and employing sorting-inspired methods such as Heapify~\cite{Zhuang_2024_setwise} and Pairwise Ranking
Prompting~\cite{qin2024largelanguagemodelseffective}.

\subsection{Adaptive Ranking Methods}
Adaptive inference methods have a long history in information retrieval, often framing document selection as a sequential decision-making problem. Early work applied multi-armed bandit algorithms to optimize ranking under click-through feedback~\cite{Li_2010}, later extending to combinatorial and semi-bandit settings that model item interactions~\cite{pmlr-v28-chen13a}. Thompson sampling~\cite{agrawal2012analysisthompsonsamplingmultiarmed} offers a Bayesian framework for balancing exploration and exploitation, with strong regret bounds in the i.i.d. setting. 

\section{Discussion}
\label{Discussion}
This work introduces \textit{contextual relevance}, a probabilistic framework for modeling document relevance as a function not only of the query but also of the surrounding batch context. Our formulation challenges standard assumptions in reranking, namely, that relevance judgments are deterministic and context independent, and instead treats them as stochastic outcomes conditioned on batch composition and ordering. Our experiments and ablations quantify the extent of contextual and intrinsic variance in LLM-based judgments and demonstrate that reranking strategies which explicitly marginalize over this variability can substantially outperform deterministic baselines under fixed inference budgets.

\section*{Limitations}
\label{limitations}
A key modeling simplification in our approach is the assumption of conditional independence across documents. While this abstraction enables tractable and effective Bayesian inference, it does not capture settings in which document-level relevance depends on interdependencies such those seen in multi-hop question answering or citation retrieval.

Extensions of our work include modeling structured dependencies between documents (e.g., via graph-based reranking or complex latent variable models) and training reranking models that produce continuous rather than binary relevance signals. These enhancements could enable more sample-efficient reranking under tight budgets and may generalize better to complex information-seeking tasks. Additionally, integrating contextual relevance into end-to-end RAG pipelines, where reranking quality directly impacts answer generation, remains a compelling area for future exploration.

\newpage
\bibliography{custom}

\newpage
\appendix

\section{Experimental Setup}
\label{app:training}

\subsection{Post-Training Setup}


We fine-tune our reranking model, Qwen2.5-7B-Instruct, over one epoch with Axolotl\footnote{\url{https://github.com/axolotl-ai-cloud/axolotl}} at a constant learning rate of 1.0e-6, global batch size 294, KL coefficient 0.01, and 7 rollouts across 42 problems per batch. For inference, we set the temperature to 0.6 for all experiments.

\subsection{Prompts}
We use the following system prompt and instructions during training and inference:
\begin{figure}[htbp]
  \centering
  \begin{minipage}{\columnwidth}
    \begin{tcolorbox}[colback=violet!5, colframe=blue!75!black, title=\textbf{\textcolor{white}{System Prompt}}, coltitle=white, fonttitle=\bfseries]
    Respond in the following format:\\
    \textless reasoning\textgreater\\
    Your detailed reasoning goes here...\\
    \textless/reasoning\textgreater\\
    \textless answer\textgreater\\
    Relevant passages: title1, title2, ...\\
    (If no passages are relevant, respond with: "Relevant passages: No relevant passages")\\
    \textless/answer\textgreater
    \end{tcolorbox}
  \end{minipage}
  \caption{System prompt used for all experiments.}
  \label{fig:system-instructions}
\end{figure}
\begin{figure}[htbp]
  \centering
  \begin{minipage}{\columnwidth}
    \begin{tcolorbox}[colback=violet!5, colframe=blue!75!black, title=\textbf{\textcolor{white}{User Instructions}}, coltitle=white, fonttitle=\bfseries]
    Identify all the relevant passages for answering the given query. Explain your reasoning step by step.
    \end{tcolorbox}
  \end{minipage}
  \caption{Instructions provided to the LLM during training and inference.}
  \label{fig:user-instructions}
\end{figure}

\section{More Experimental Results}
\label{app:more_experimental_results}
\subsection{BRIGHT Results.}
\label{app:bright_results}

Table~\ref{tab:main_results_app} reports nDCG@10 results on BRIGHT, split by topic with an inference budget of \(T=100\).

\begin{table*}[t]
\centering
\footnotesize
\resizebox{\textwidth}{!}{%
\begin{tabular}{lcccccccc}
\toprule
\textbf{Topic} & \textbf{BM25s} & \textbf{Heapify} & \textbf{Uniform} & \textbf{TS-75/25} & \textbf{TS-50/50} & \textbf{TS-25/75} & \textbf{TS-0/100} \\
\midrule
AoPS                & 0.01582 & 0.0158 & 0.017\sd{0.003} & 0.013\sd{0.001} & 0.016\sd{0.003} & 0.014\sd{0.004} & 0.011\sd{0.001} \\
Biology             & 0.44682 & 0.4800 & 0.480\sd{0.007} & 0.497\sd{0.006} & 0.493\sd{0.015} & 0.480\sd{0.006} & 0.489\sd{0.012} \\
Earth Science       & 0.52155 & 0.5140 & 0.459\sd{0.013} & 0.462\sd{0.015} & 0.458\sd{0.007} & 0.466\sd{0.017} & 0.453\sd{0.004} \\
Economics           & 0.27620 & 0.3030 & 0.276\sd{0.005} & 0.300\sd{0.006} & 0.289\sd{0.014} & 0.284\sd{0.009} & 0.294\sd{0.008} \\
Psychology          & 0.36296 & 0.4050 & 0.439\sd{0.008} & 0.468\sd{0.004} & 0.464\sd{0.005} & 0.476\sd{0.006} & 0.463\sd{0.003} \\
Robotics            & 0.16064 & 0.2040 & 0.281\sd{0.013} & 0.291\sd{0.007} & 0.290\sd{0.009} & 0.292\sd{0.003} & 0.286\sd{0.010} \\
Stack Overflow      & 0.27077 & 0.2760 & 0.275\sd{0.006} & 0.292\sd{0.009} & 0.294\sd{0.012} & 0.299\sd{0.013} & 0.297\sd{0.003} \\
Sustainable Living  & 0.23136 & 0.2820 & 0.373\sd{0.012} & 0.384\sd{0.010} & 0.390\sd{0.010} & 0.383\sd{0.004} & 0.374\sd{0.002} \\
Leetcode            & 0.12262 & 0.1160 & 0.164\sd{0.008} & 0.144\sd{0.007} & 0.138\sd{0.014} & 0.129\sd{0.003} & 0.119\sd{0.005} \\
Pony                & 0.08560 & 0.0879 & 0.189\sd{0.002} & 0.201\sd{0.002} & 0.199\sd{0.004} & 0.200\sd{0.003} & 0.199\sd{0.003} \\
TheoremQA-Q         & 0.11721 & 0.1150 & 0.127\sd{0.002} & 0.117\sd{0.004} & 0.113\sd{0.003} & 0.104\sd{0.003} & 0.086\sd{0.004} \\
TheoremQA-T         & 0.20301 & 0.2720 & 0.368\sd{0.014} & 0.360\sd{0.011} & 0.369\sd{0.004} & 0.353\sd{0.004} & 0.363\sd{0.010} \\
\midrule
\textbf{All topics} & 0.23455 & 0.2560 & 0.287\sd{0.003} & 0.294\sd{0.002} & 0.293\sd{0.003} & 0.290\sd{0.004} & 0.286\sd{0.002} \\
\bottomrule
\end{tabular}%
}
\caption{nDCG@10 performance under \(T=100\) on BRIGHT split by topic.}
\label{tab:main_results_app}
\end{table*}

\subsection{BEIR Results.}

Table~\ref{tab:beir_results} reports nDCG@10 results on BEIR datasets with fewer than 2,000 queries. We sample 100 queries per dataset.

\label{app:beir_results}
\begin{table*}[t]
\centering
\footnotesize
\resizebox{\textwidth}{!}{%
\begin{tabular}{lcccccccc}
\toprule
\textbf{Dataset}       & \textbf{BM25s} & \textbf{Heapify}   & \textbf{Uniform}       & \textbf{TS-75/25}      & \textbf{TS-50/50}      & \textbf{TS-25/75}      & \textbf{TS-0/100}      \\
\midrule
ArguAna               & 0.2850 & 0.4700 & 0.4850\sd{0.0190}     & 0.4640\sd{0.0020}     & 0.4530\sd{0.0230}     & 0.4620\sd{0.0080}     & 0.4710\sd{0.0150}     \\
Climate-FEVER         & 0.1410 & 0.1440 & 0.1840\sd{0.0040}     & 0.1990\sd{0.0030}     & 0.2010\sd{0.0090}     & 0.2030\sd{0.0070}     & 0.2090\sd{0.0140}     \\
DBPedia               & 0.3130 & 0.3300 & 0.3580\sd{0.0030}     & 0.3680\sd{0.0030}     & 0.3710\sd{0.0080}     & 0.3820\sd{0.0060}     & 0.3810\sd{0.0030}     \\
FiQA-2018             & 0.2320 & 0.3090 & 0.3730\sd{0.0080}     & 0.4030\sd{0.0010}     & 0.4020\sd{0.0080}     & 0.4030\sd{0.0050}     & 0.4030\sd{0.0100}     \\
NFCorpus              & 0.3560 & 0.3820 & 0.4040\sd{0.0030}     & 0.4010\sd{0.0040}     & 0.3980\sd{0.0040}     & 0.3990\sd{0.0040}     & 0.3950\sd{0.0010}     \\
SCIDOCS               & 0.1320 & 0.1380 & 0.1570\sd{0.0050}     & 0.1590\sd{0.0010}     & 0.1550\sd{0.0020}     & 0.1630\sd{0.0020}     & 0.1480\sd{0.0110}     \\
SciFact               & 0.7200 & 0.7490 & 0.7330\sd{0.0210}     & 0.7550\sd{0.0130}     & 0.7460\sd{0.0140}     & 0.7540\sd{0.0500}     & 0.7430\sd{0.0200}     \\
TREC-COVID            & 0.5930 & 0.6950 & 0.8060\sd{0.0060}     & 0.8210\sd{0.0090}     & 0.8110\sd{0.0060}     & 0.8200\sd{0.0100}     & 0.8110\sd{0.0120}     \\
Touche-2020           & 0.4410 & 0.4570 & 0.2910\sd{0.0250}     & 0.2940\sd{0.0190}     & 0.2820\sd{0.0110}     & 0.2950\sd{0.0250}     & 0.2940\sd{0.0090}     \\
\midrule
\textbf{Average}      & 0.3570
                      & 0.4080
                      & 0.421\sd{0.003}
                      & 0.429\sd{0.002}
                      & 0.424\sd{0.004}
                      & 0.431\sd{0.010}
                      & 0.428\sd{0.005} \\
\bottomrule
\end{tabular}%
}
\caption{nDCG@10 performance under \(T=100\) on BEIR split by dataset.}
\label{tab:beir_results}
\end{table*}

\section{TS-SetRank-T: Throughput-Optimized Variant}
\label{app:ts_setrank_t}
While the TS-SetRank algorithm (Algorithm~\ref{alg:ts-setrank}) in the main paper outlines our general two-phase Bayesian inference strategy, we also introduce TS-SetRank-T, a throughput-optimized variant designed for environments with batched or delayed LLM calls to increase the parallelism in Phase II.

Unlike TS-SetRank, which performs posterior updates immediately after each batch, TS-SetRank-T (Algorithm~\ref{alg:ts-setrank-t}) aggregates binary feedback across multiple rounds before updating. The update interval $\tau$ controls how frequently the posterior is updated, effectively governing the number of queries that can be executed concurrently in Phase II. In Phase I, all inference rounds are trivially parallelizable due to the uniform sampling strategy.

\begin{algorithm}[t]
\caption{TS-SetRank-T (Throughput-Optimized)}
\KwIn{First-stage retrieval results $D = \{d_1, \dots, d_N\}$, query $q$, budget $T$, batch size $b$, exploration rounds $T_f$, update interval $\tau$, reranking model $M$}
\KwOut{Final reranked list of documents}

\ForEach{$d_i \in D$}{
  Initialize posterior: $\alpha_i \gets 1$, $\beta_i \gets 1$ \\
  Initialize counters: $S_i \gets 0$, $F_i \gets 0$
}

\For{$t = 1$ \KwTo $T$}{
  \uIf{$t \leq T_f$}{
    Sample batch $S_t \sim \mathrm{Uniform}\!\bigl(D_b(D)\bigr)$
  }
  \Else{
    \ForEach{$d_i \in D$}{
      Draw $\tilde\theta_i \sim \mathrm{Beta}(\alpha_i, \beta_i)$
    }
    Let $S_t \gets \{\,d_i \in D : \tilde\theta_i \text{ is among the top-}b \,\}$
  }

  Obtain feedback: $R_t := R(q, S_t)$ from reranking model $M$

  \ForEach{$d_i \in S_t$}{
    \uIf{$d_i \in R_t$}{ $S_i \gets S_i + 1$ }
    \Else{ $F_i \gets F_i + 1$ }
  }

  \If{$t > T_f$ \textbf{and} $(t - T_f) \bmod \tau = 0$}{
    \ForEach{$d_i \in D$}{
      $\alpha_i \gets \alpha_i + S_i$, \quad $\beta_i \gets \beta_i + F_i$ \\
      $S_i \gets 0$, \quad $F_i \gets 0$
    }
  }
}

\ForEach{$d_i \in D$}{
  Final update: $\alpha_i \gets \alpha_i + S_i$, \quad $\beta_i \gets \beta_i + F_i$
}

\Return documents sorted by posterior means $\hat\theta_i = \alpha_i / (\alpha_i + \beta_i)$ in descending order.
\label{alg:ts-setrank-t}
\end{algorithm}

\end{document}